\documentclass[preprintnumbers,nofootinbib,a4paper]{revtex4-1}

\usepackage{amsmath,latexsym,amssymb,amsfonts}
\usepackage{float} \usepackage{graphicx}
\usepackage{epstopdf}
\usepackage{graphics}
\usepackage{caption}
\usepackage{subfigure}
\usepackage{bm}
\usepackage{color}
\usepackage{hyperref}
\usepackage[dvips]{psfrag}

\begin{document}

\title{Inflation with Derivative Self-interaction and Coupling to Gravity}

\author{Gansukh Tumurtushaa}
\email[E-mail: ]{gansuhmgl@ibs.re.kr}
\affiliation{Center for Theoretical Physics of the Universe, Institute for Basic Science (IBS), Daejeon 34051, Korea}
\preprint{CTPU-PTC-19-09}
\begin{abstract}
We consider a subclass of Horndeski theories for studying cosmic inflation. In particular, we investigate models of inflation in which the derivative self-interaction of the scalar field and the non-minimal derivative coupling to gravity are present in the action and equally important during inflation. In order to control contributions of each term as well as to approach the single-term limit, we introduce a special relation between the derivative interaction and the coupling to gravity. By calculating observable quantities including the power spectra and spectral tilts of scalar and tensor perturbation modes, and the tensor-to-scalar ratio, we found that the tensor-to-scalar ratio is suppressed by a factor of $(1+1/\gamma)$, where $\gamma$ reflects the strength of the derivative self-interaction of the inflaton field with respect to the derivative coupling gravity. We placed observational constraints on the chaotic and natural inflation models and showed that the models are consistent with the current observational data mainly due to the suppressed tensor-to-scalar ratio.
\end{abstract}


\maketitle

\section{Introduction}

Inflation in the early universe is a successful paradigm for explaining the cosmological problems including the horizon, flatness and monopole problems~\cite{Guth:1980zm, Linde:1981mu, Albrecht:1982wi, Sato:1980yn}. Driven by a slowly rolling scalar field (or "inflaton"), inflation also generates the primordial density perturbations necessary for the formation of large scale structures in the universe~\cite{Mukhanov:1981xt, Hawking:1982cz, Starobinsky:1982ee, Guth:1982ec, Bardeen:1983qw, Kodama:1985bj}. 
The constraints on inflationary models, in particular, sufficient amount of inflation and cosmic microwave background (CMB) temperature anisotropy measurements favor a nearly flat potential during inflation~\cite{Akrami:2018odb}. In addition to the nearly flat potential, the conventional scalar-field action also consists of a canonical kinetic term~\cite{Yamaguchi:2011kg, Martin:2013tda}. Arising naturally from particle physics, inflationary models with non-canonical kinetic terms have received much attention over the past few decades as they can reconcile the simplest realization of inflation with the current observational data and leave their distinct signatures in the cosmological observations~\cite{ArmendarizPicon:1999rj, Kobayashi:2010cm, Silverstein:2003hf, Alishahiha:2004eh, Fujii:1982ms}. 

It is demonstrated that special combinations of higher-order kinetic terms in the action give rise to the equations of motion that contain no higher than the second-order derivatives~\cite{Nicolis:2008in, Deffayet:2009wt}. Applications of such extended scenarios provide a unified framework upon which one can construct or embed new models of inflation. The Horndeski theories as a generalization -- or an extension -- of the scalar-tensor theories of gravity have the most general higher derivative extensions and its dynamics is governed by second-order equations of motion~\cite{Horndeski:1974wa}. According to Refs.~\cite{Deffayet:2011gz, Kobayashi:2011nu}, the most general four-dimensional scalar-tensor theories possessing second-order equations of motion are described by the following action
\begin{eqnarray}\label{eq:Horndenski}
S = \int d^{4}x \left( L_2+L_3+L_4+L_5 \right)\,, 
\end{eqnarray}
with 
{\small \begin{eqnarray} \label{eq:Lag}
L_2 &=& K(\phi, X)\,,\nonumber\\ \quad L_3 &=& G_3(\phi, X) \square\phi\,,\nonumber\\ \quad L_4 &=& G_4(\phi, X) R + G_{4,X}\left[\left(\square \phi\right)^2 - \left(\nabla_\mu\phi\nabla_\nu\phi\right)\left(\nabla^\mu\phi\nabla^\nu\phi\right)\right]\,,\\
L_5 &=& G_5(\phi, X) G_{\mu\nu}\left(\nabla^\mu\phi\nabla^\nu\phi\right)-\frac16 G_{5,X}\left[(\square\phi)^3 -3 \square\phi \left(\nabla_\mu\phi\nabla_\nu\phi\right)\left(\nabla^\mu\phi\nabla^\nu\phi\right)+2 \left(\nabla^\mu\phi\nabla_\alpha\phi\right)\left(\nabla^\alpha\phi\nabla_\beta\phi\right)\left(\nabla^\beta\phi\nabla_\mu\phi\right)\right]\,,\nonumber
\end{eqnarray}}
where $X = -\nabla_\mu\phi\nabla^\mu\phi/2$ and $\square\phi = \nabla_\mu\nabla^\mu\phi$. Here, $K(\phi, X)$ and $G_{3,4,5}(\phi, X)$ are arbitrary functions of the scalar field $\phi$ and $X$ and $G_{i,X}(\phi,X) = \partial G(\phi,X)/\partial X$ with $i=4,5$. In Ref.~\cite{ Kobayashi:2011nu}, Lagrangians in Eq.~(\ref{eq:Lag}) are shown to be equivalent to the ones discovered by Horndeski. The action reduces to the general relativity if $K=G_3=G_5=0$ and $G_4=1/2\kappa^2$, where $\kappa=1/M_{pl}$ is the inverse of the reduced Planck mass $M_{pl}$. 

A broad spectrum of single-field models of inflation with the second-order equations of motion are constructed from Eq.~(\ref{eq:Horndenski}) and the associated cosmological perturbations are well established in Refs.~\cite{Kobayashi:2011nu, Gao:2012ib}. Thus far, the successful proposals of inflationary models within this framework often employ special combinations of the independent functions of the scalar field and its derivatives: $K(\phi, X)$ and $G_{3,4,5}(\phi, X)$. This is mainly due to the fact that it is nontrivial how the background as well as perturbations evolve when multiple terms are present in the action. For example, the inflationary models with the non-minimal derivative coupling to gravity~\cite{Germani:2010gm, Tsujikawa:2012mk, Yang:2015zgh, Sato:2017qau, Amendola:1993uh} have focused on a case where all terms except the $L_3$ are present in the action. On the other hand, the inflationary models with the derivative self-interactions of the scalar field (\emph{i.e.,} G-inflation)~\cite{Kobayashi:2010cm, Kobayashi:2011nu,  Kobayashi:2011pc, Kamada:2010qe} (see Ref.~\cite{Kamada:2010qe} for the potential driven G-inflation scenarios) concentrate on the presence of $L_3$ by omitting $L_5$ in Eq.~(\ref{eq:Horndenski}). 
However, although the general formulas are well established in Refs.~\cite{Kobayashi:2011nu, Gao:2012ib}, little attention has been paid to the case where all terms in Eq.~(\ref{eq:Horndenski}) are present in the action and equally important during inflation. 

Our aim for the present work is to investigate the potential driven single-field models of slow-roll inflation in which all terms in Eq.~(\ref{eq:Horndenski}) are present and equally important during inflation. In addition, by introducing a special relation, we show how they could approach the single term limit with respect to model parameters. 
We focus in particular on the following setup: 
\begin{eqnarray}\label{eq:setup}
K(\phi, X) = X- V\,, \qquad G_{3}(\phi,X) = -c_3 \xi(\phi) X \,, \qquad G_4(\phi, X) = \frac{1}{2 \kappa^2}\,, \qquad G_5=-\frac{1}{2}c_1\phi\,,
\end{eqnarray}  
where $V(\phi)$ is the inflaton potential, $c_1$ and $c_3$ are the model parameters, and $\xi(\phi)$ is the coupling function of $\phi$. In fact, aforementioned two classes of inflationary models; namely, inflation with the derivative self-interaction of the scalar field and inflation with the non-minimal derivative coupling to gravity, are combined into one setup if $c_1\neq0$ and $c_3\neq0$ in Eq.~(\ref{eq:setup}). We examine the observational consistency of chaotic inflation with $V(\phi)=\lambda (\kappa \phi)^n/(n \kappa^4)$ and natural inflation with $V(\phi)=\Lambda^4\left[1+\cos(\phi/f)\right]$ for our setup Eq.~(\ref{eq:setup}). This is because theoretical predictions of these models for standard single-field inflation with Einstein gravity are disfavored by the current observational data~\cite{Akrami:2018odb}.

This paper is organized as follows: in Sec.~\ref{sec:setup}, after deriving the background equations of motion for our setup, we introduce a special relation between the $L_3$ and $L_5$ of Eq.~(\ref{eq:Horndenski}) that allows us to control contributions of each term during inflation. In Sec.~\ref{sec:pert}, following Ref.~\cite{Kobayashi:2011nu}, we obtain the observable quantities including the power spectra ($\mathcal{P}_S$, $\mathcal{P}_T$) and spectral tilts of the scalar and tensor perturbation modes ($n_S$, $n_T$), and the tensor-to-scalar ratio $r$. Our results of observational constraints on the chaotic and natural inflation models are presented in Sec.~\ref{sec:ObsCons} and we conclude our work in Sec.~\ref{sec:conc}. 

\section{Setup for potential-driven slow-roll inflation}\label{sec:setup}
For our setup given in Eq.~(\ref{eq:setup}), the action Eq.~(\ref{eq:Horndenski}) reduces to
\begin{eqnarray}\label{eq:NMDCandGinf}
S = \int d^{4}x\sqrt{-g} \left[ \frac{1}{2\kappa^{2}} R - \frac{1}{2} \left(g^{\mu\nu}+c_1 G^{\mu\nu}\right)\partial_\mu\phi\partial_\nu \phi - V(\phi) - \frac{1}{2}c_3\xi(\phi)\partial_\mu\phi\partial^\mu\phi\partial_\nu\partial^\nu\phi \right]\,,
\end{eqnarray}
where $c_1\neq0$ and $c_3\neq0$ constants. Varying this action with respect to metric $g_{\mu\nu}$ yields the Einstein equation 
\begin{eqnarray}\label{eq:Einsteineq}
&&G_{\mu\nu}= \kappa^2 \left(
T_{\mu\nu}^{\phi}+ c_1 T_{\mu\nu}^{c_1} + c_3 T_{\mu\nu}^{c_3}\right)\,,
\end{eqnarray}
where the energy-momentum tensors $T_{\mu\nu}^{\phi}$, $T_{\mu\nu}^{c_1}$, and $T_{\mu\nu}^{c_3}$ are given by
\begin{eqnarray}
T_{\mu\nu}^{\phi} &=& \partial_\mu\phi \partial_\nu\phi-\frac{1}{2}g_{\mu\nu}\left(\partial_\alpha\phi\partial^\alpha\phi+2V\right)\,,\\
T_{\mu\nu}^{c_1} &=& -\frac{1}{2}\nabla_\mu\phi \nabla_\nu\phi R + 2\nabla_\alpha\phi \nabla_{(\mu}\phi R^{\alpha}\,_{\nu)}+\nabla^\alpha\phi\nabla^\beta\phi R_{\mu\alpha\nu\beta}+\nabla_\mu\nabla^\alpha\phi\nabla_\nu\nabla_\alpha\phi\nonumber\\
& &-\nabla_\mu\nabla_\nu\phi\square\phi-\frac12 G_{\mu\nu} \nabla_\alpha\phi \nabla^\alpha\phi+g_{\mu\nu}\left[-\frac{1}{2} \nabla^\alpha\nabla^\beta\phi\nabla_\alpha\nabla_\beta\phi+\frac{1}{2}\left(\square\phi\right)^2-\nabla_\alpha\phi\nabla_\beta\phi R^{\alpha\beta}\right]\,,\\
T_{\mu\nu}^{c_3}&=& \frac{1}{2}\left[( \xi \nabla_\alpha\phi\nabla^\alpha\phi)_{(\mu}\nabla_{\nu)}\phi - \xi\square\phi \nabla_\mu\phi\nabla_\nu\phi-\frac{1}{2}g_{\mu\nu}(\xi\nabla_\alpha\phi\nabla^\alpha\phi)_{\beta}\nabla^\beta\phi\right] \,,
\end{eqnarray}
respectively. Using the Bianchi identity $\nabla^\mu G_{\mu\nu}=0$ and the conservation law $\nabla^\mu T_{\mu\nu}=0$, we get from Eq.~(\ref{eq:Einsteineq}) 
\begin{eqnarray}
\nabla^{\mu} \left(T^\phi_{\mu\nu}+c_1 T_{\mu\nu}^{c_1} +c_3 T_{\mu\nu}^{c_3} \right) = 0\,, 
\end{eqnarray}
which, as a consequence, yields a evolution equation for the scalar field.

In a spatially flat Friedman-Robertson-Walker universe with metric 
\begin{eqnarray}\label{eq:flatMetric}
ds^2=-dt^2+a(t)^2\delta_{ij}dx^i dx^j\,,
\end{eqnarray} 
where $a(t)$ is a scale factor, the background Einstein and field equations are obtained as
\begin{eqnarray}
&& H^2 = \frac{\kappa^2}{3}\left[\frac{1}{2} \dot{\phi}^2 +V(\phi) -\frac{9}{2}c_1\dot{\phi}^2 H^2 +\frac{1}{2}c_3\dot{\phi}^3 \left( \dot{\xi}-6H\xi\right) 
\right]\,,\label{eq:EE00}\\
&& 2\dot{H}+3H^2 = -\frac{\kappa^2}{2}\left[ \dot{\phi}^2 -2V +c_1\dot{\phi}^2\left(2\dot{H} +3H^2 +4H\frac{\ddot{\phi}}{\dot{\phi}} \right) +c_3\dot{\phi}^2\left(2\xi \ddot{\phi} + \dot{\xi}\dot{\phi}\right) 
\right]\,,\\
&& \ddot{\phi} +3H\dot{\phi} + V_{,\phi}- 3 c_1 H \dot{\phi}\left( 2\dot{H} +3H^2 +H\frac{\ddot{\phi}}{\dot{\phi}}\right)+\frac{1}{2}c_3\dot{\phi}\left[\ddot{\xi}\dot{\phi}+3\dot{\xi}\ddot{\phi} - 6\xi\dot{\phi} \left( \dot{H} + 3H^2 + 2 H \frac{\ddot{\phi}}{\dot{\phi}} \right) \right]=0\,.\label{eq:fieldEq}
\end{eqnarray}

For the slow-roll inflation, we introduce so called the slow-roll conditions that read as $V(\phi)\gg \dot{\phi}^2$ and $\ddot{\phi}\ll 3H\dot{\phi}$. In order to quantify these slow-roll conditions, it is useful to introduce the slow-roll parameters~\cite{Tsujikawa:2012mk, Kobayashi:2010cm}
\begin{eqnarray}\label{eq:srparams}
\epsilon_1 \equiv -\frac{\dot{H}}{H^2}\,,\quad 
\epsilon_2 \equiv -\frac{\ddot{\phi}}{H\dot{\phi}}\,,\quad 
\epsilon_3 \equiv\frac{\xi_{,\phi}\dot{\phi}}{\xi H}\,,\quad 
\epsilon_4 \equiv \frac{\xi_{,\phi\phi}\dot{\phi}^4}{V_{,\phi}}\,, \quad 
\epsilon_5 \equiv \frac{\kappa^2\dot{\phi}^2}{2H^2}\,, 
\end{eqnarray}
which assumed to be small during inflation. Thus, Eq.~(\ref{eq:fieldEq}) can be rewritten in terms of these parameters as
\begin{eqnarray}\label{eq:fieldEqSR}
3H\dot{\phi}\left[1-\frac{1}{3}\epsilon_2 -c_1 H^2(3-2\epsilon_1-\epsilon_2) -c_3\xi H\dot{\phi}\left( 3-\epsilon_1-2\epsilon_2-\frac{2}{3}\epsilon_2\epsilon_3\right) \right] = -V_{,\phi}\left(1-\frac{1}{2}c_3 \epsilon_4 \right)\,.
\end{eqnarray}
In addition to usual slow-roll conditions, we introduce the following relation between $L_3$ and $L_5$ in Eq.~(\ref{eq:Horndenski}):
\begin{eqnarray}\label{eq:Addcond}
c_1 H^2=\gamma c_3\xi\dot{\phi}H \,,
\end{eqnarray} 
where $\gamma$ is a constant reflecting the strength of the inflaton derivative self-interaction ($L_3$) with respect to the the non-minimal derivative coupling to gravity ($L_5$). The effects of $L_5$ term dominates over that of $L_3$ when $\gamma\gg 1$ and vice versa when $\gamma\ll1$. The both terms are equally important during inflation when $\gamma\sim\mathcal{O}(1)$, which we are more interested in this work study. Although it is possible to find a set of $\xi(\phi)$ and $V(\phi)$ that fits well to the observational data without introducing this relation, introducing Eq.~(\ref{eq:Addcond}) allows us to control the contributions of each term through the $\gamma$ parameter. A noteworthy feature of this relation is that the shape of $\xi(\phi)$ during inflation can be determined by  Eq.~(\ref{eq:Addcond}) once $V(\phi)$ is known. 

Under the slow-roll conditions, the terms containing $c_1$ and $c_3$ in Eq.~(\ref{eq:EE00}) are much  smaller than $V(\phi)$ during inflation hence Eqs.~(\ref{eq:EE00}) and (\ref{eq:fieldEqSR}) reduce to
\begin{eqnarray}\label{eq:srapproxEq}
3H^2\simeq \kappa^2 V\,, \quad 3H\dot{\phi} \left( 1+\mathcal{A}\right) \simeq -V_{,\phi} \,,
\end{eqnarray} 
where $\mathcal{A}
\equiv-3 c_1 (1+1/\gamma) H^2$ after taking Eq.~(\ref{eq:Addcond}) into account. When $\mathcal{A}\gg1$, the friction  term significantly enhances hence it is regarded as the high friction limit, see Refs.~\cite{Germani:2010gm, Tsujikawa:2012mk, Yang:2015zgh, Sato:2017qau} for the further details. On the other hand, the standard slow-roll inflation with Einstein gravity is realized when $|\mathcal{A}|\ll 1$. Thus, terms with $c_1$ and $c_3$ play an important role when $|\mathcal{A}|\gg 1$. 
We derive potential based slow-roll parameters using Eq.~(\ref{eq:srapproxEq})
\begin{eqnarray}\label{eq:potSR}
\epsilon_1&=&\frac{\epsilon_V}{1+\mathcal{A}}\,,\quad 
\epsilon_2
\simeq\frac{\eta_V-3\epsilon_V}{1+\mathcal{A}}+\frac{2\epsilon_V}{(1+\mathcal{A})^2}\,,\quad 
\epsilon_3
\simeq\frac{\eta_V-4\epsilon_V}{1+\mathcal{A}}+\frac{2\epsilon_V}{(1+\mathcal{A})^2}\,,
\end{eqnarray}
where
\begin{eqnarray}
\epsilon_V &=& \frac{1}{2\kappa^2}\left(\frac{V_{,\phi}}{V}\right)^2\,,\qquad \eta_V=\frac{V_{,\phi\phi}}{\kappa^2 V}\,.
\end{eqnarray}
The amount of inflation is quantified by the number $N$ of $e$-folds, which reads
\begin{eqnarray}\label{eq:Nofefold}
N 
=\int^{\phi_e}_\phi \frac{H}{\dot{\phi}}d\phi' 
\simeq \kappa^2\int_{\phi_e}^\phi \frac{ V}{V_{,\phi'}}(1+\mathcal{A}) d\phi'\,,
\end{eqnarray}
where $\phi_e$ is the scalar-field value at the end of inflation and is to be estimated by solving $\epsilon_1(\phi_e)\equiv 1$. 

\section{Linear Perturbation Analysis}\label{sec:pert}
In this section, we discuss the linear perturbation analyses for scalar and tensor modes in the flat FRW background and our discussion mainly follows Ref.~\citep{Kobayashi:2011nu} as the most general perturbation analysis for the Horndeski theories is carried out there in great detail. The perturbed metric in the Arnowitt-Deser-Misner formalism~\cite{ADMformalism} is given by
\begin{eqnarray}\label{eq:pertgenMet}
ds^2=-N^2dt^2 + \gamma_{ij}(dx^i+N^idt)(dx^j+N^jdt)\,,
\end{eqnarray}
where $N$, $N^i$, and $\gamma_{ij}$ are the lapse function, the shift function, and the metric for the three-dimensional space, respectively, and are given by 
\begin{eqnarray}\label{eq:metricPert}
N=1+\alpha\,,\quad N_i=\partial_i\beta\,,\quad  \gamma_{ij}=a^2(t)e^{2\zeta}\left(\delta_{ij}+h_{ij}+\frac{1}{2}h_{ik}h_{kj}\right)\,.
\end{eqnarray}
Here, $\alpha$, $\beta$, and $\zeta$ denote scalar perturbations while $h_{ij}$ is a tensor perturbation satisfying the traceless and transverse conditions; $h_{ii}=0=h_{ij,j}$. The scalar field is decomposed into a background and inhomogeneous parts, \emph{e.g.,} $\phi(t,\bold{x})=\phi(t)+\delta\phi(t,\bold{x})$, and we employ the uniform field gauge with $\delta\phi(t,\bold{x})=0$.

\subsection{Tensor perturbations}
Let us first consider the tensor perturbations. Substituting the perturbed metric into the action Eq.~(\ref{eq:NMDCandGinf}) and then expanding the action to the second order in $h_{ij}$, one can obtain the quadratic action as~\citep{Kobayashi:2011nu}
\begin{eqnarray}\label{eq:quadAction}
S_T^{(2)} = \frac{1}{8}\int dt d^3x a^3 \left[ G_T \dot{h}_{ij}^2-\frac{1}{a^2}F_T(\partial_k h_{ij})^2 \right]\,,
\end{eqnarray} 
where 
\begin{eqnarray}
G_T = \frac{1}{\kappa^2}\left(1 +\frac{1}{2}c_1\kappa^2\dot{\phi}^2\right)\,,\quad
F_T = \frac{1}{\kappa^2}\left(1 -\frac{1}{2}c_1\kappa^2\dot{\phi}^2\right)\,.
\end{eqnarray}
In order to avoid from the ghost and gradient instabilities, the $G_T>0$ and $F_T>0$ conditions must be satisfied in Eq.~(\ref{eq:quadAction}). In terms of slow-roll parameters, we get
\begin{eqnarray}
G_T 
\simeq \frac{1}{\kappa^2} \left[1-\frac{\epsilon_V}{3(1+1/\gamma)(1+\mathcal{A})}\right]>0\,,\quad 
F_T 
\simeq \frac{1}{\kappa^2} \left[1+\frac{\epsilon_V}{3(1+1/\gamma)(1+\mathcal{A})}\right]>0\,,
\end{eqnarray}
By decomposing the tensor perturbation 
$
h_{ij}=\sum_{\lambda=+,\times} \epsilon^\lambda_{ij}h_\lambda\,,
$
where $\epsilon_{ij}$ is a polarization tensor satisfying $\sum_{i}\epsilon_{ii}^\lambda=0$ and $\sum_{i, j}\epsilon_{ij}^\lambda\epsilon_{ij}^{\lambda'}=\delta^{\lambda\lambda'}$, and using the canonical variable 
$
u_{\lambda}=z_T h_\lambda\,,
$
where
$
z_T =a\left(G_T F_T\right)^{1/4}/2\,,
$ 
the quadratic action Eq.~(\ref{eq:quadAction}) is rewritten as
\begin{eqnarray}
S_T^{(2)} =\frac{1}{2} \sum_{\lambda= +,\times}\int d\tau d^3x\left[ (u_\lambda')^2-c_T^2(\partial_k u_\lambda)^2+\frac{z_T''}{z_T}u_\lambda^2\right]\,,
\end{eqnarray}
where 
\begin{eqnarray}\label{eq:soundSpeed}
c_T^2&=&\frac{F_T}{G_T}
\simeq 1+\frac{2\epsilon_V}{3(1+1/\gamma)(1+\mathcal{A})}\,,\\
\frac{z''_{T}}{z_T}&\simeq & 
\frac{1}{\tau^2}\left(2+\frac{3\epsilon_V}{1+\mathcal{A}}\right)\,.
\end{eqnarray}
Here, the prime denotes the derivative with respect to the conformal time $\tau$ which relates the physical time $t$ via $ad\tau=c_T dt$. 
Each perturbation mode crosses the sound horizon when $k^2=z_T''/z_T\sim1/\tau^2$, where $k$ is the wavenumber. 

By employing the canonical quantization
\begin{eqnarray}
\hat{u}_\lambda = \int \frac{d^3k}{(2\pi)^3}\left[u_{\lambda, k}(\tau)\hat{a}_k e^{i\vec{k}\vec{x}}+u^\ast_{\lambda,k}(\tau)\hat{a}^\dagger e^{-i\vec{k}\vec{x}} \right]\,,
\end{eqnarray} 
we obtain a wave equation for the tensor perturbation modes
\begin{eqnarray}\label{eq:TmodeEq}
u_{\lambda,k}''+\left(c_T^2k^2-\frac{\mu_T^2-1/4}{\tau^2}\right)u_{\lambda,k}=0\,,
\end{eqnarray}
where
\begin{eqnarray}
\frac{z''_T}{z_T}=\frac{\mu_T^2-1/4}{\tau^2}\,,\quad \mu_T \simeq 
\frac{3}{2}+\frac{\epsilon_V}{1+\mathcal{A}}\,.
\end{eqnarray} 
The exact solution to Eq.~(\ref{eq:TmodeEq}) can be obtained by adopting the Bunch-Davies vacuum for the initial fluctuation modes at $c_T k |\tau|\gg 1$ and assuming constant slow-roll parameters during inflation. The solution therefore reads
\begin{eqnarray}\label{eq:TensSol}
u_{\lambda,k} = 2^{\mu_T-\frac{3}{2}}\frac{\Gamma(\mu_T)}{\Gamma(3/2)}\frac{e^{i(\mu_T-\frac{1}{2})\frac{\pi}{2}}}{\sqrt{2c_T k}}\left(-c_Tk\tau\right)^{\frac{1}{2}-\mu_T}\,.
\end{eqnarray}
The power spectra of the tensor modes can be calculated with Eq.~(\ref{eq:TensSol}) on the large scale $c_Tk|\tau|\ll 1$ as
\begin{eqnarray}\label{eq:Pt}
\mathcal{P}_T&=&\frac{k^3}{\pi^2}\sum_{\lambda=+, \times}\left|\frac{u_{\lambda, k}}{z_T}\right|^2
\simeq \frac{\kappa^2 H^2}{2\pi^2c_T^3}\,.
\end{eqnarray}
The tensor spectral index at the time of horizon crossing is computed as
\begin{eqnarray}\label{eq:ntindex}
n_T=\left.\frac{\ln \mathcal{P}_T}{\ln k}\right|_{c_T k = a H}=3-2\mu_T\simeq-\frac{2\epsilon_V}{1+\mathcal{A}}\,.
\end{eqnarray}

\subsection{Scalar perturbations}
Next, let us discuss the scalar perturbations by setting $h_{ij}=0$ in Eq.~(\ref{eq:metricPert}). Substituting the perturbed metric into the action in  Eq.~(\ref{eq:NMDCandGinf}) and then expanding to the second order in $\zeta$, one can also obtain the action~\citep{Kobayashi:2011nu}
\begin{eqnarray}\label{eq:ScalPertAct}
S_S^{(2)}=\frac{1}{\kappa^2}\int dt d^3x a^3\left[G_S\dot{\zeta}^2-\frac{1}{a^2}F_S(\partial_i\zeta)^2 \right]\,,
\end{eqnarray}
where
\begin{eqnarray}
G_S&=&\frac{\Sigma}{\Theta^2} G_T^2+3G_T\,,\\
F_S&=&\frac{1}{a}\frac{d}{dt}\left(\frac{a}{\Theta}G_T^2\right)-F_T = \frac{G_T^2}{\Theta}H\left(1+\frac{2\dot{G_T}}{G_T H}-\frac{\dot{\Theta}}{\Theta H} -\frac{\Theta F_T}{G_T^2 H} \right)\,,
\end{eqnarray}
with
\begin{eqnarray}\label{eq:Sigma}
\Sigma &=& \frac{1}{2}\dot{\phi}^2-\frac{3}{\kappa^2}H^2-9c_1H^2\dot{\phi}^2-c_3 H \xi \dot{\phi}^3\left(6-\frac{\dot{\xi}}{\xi H}\right)\,,\\ 
\Theta &=& \frac{H}{\kappa^2}\left(1+\frac{3}{2}c_1\kappa^2\dot{\phi}^2 \right)+\frac{1}{2}c_3\xi \dot{\phi}^3\,.
\label{eq:Theta}
\end{eqnarray}
Here, $F_S>0$ and $G_S>0$ are also necessary for avoiding the ghost and gradient instabilities.
By introducing the canonically normalized field $v=z_S \zeta$ with $z_S=\sqrt{2}a\left(G_S F_S\right)^{1/4}$, one can rewrite Eq.~(\ref{eq:ScalPertAct}) as
\begin{eqnarray}
S_S^{(2)}=\frac{1}{2}\int d\tau d^3x \left[v'^2-c_S^2(\partial_i v)^2 +\frac{z_S''}{z_S}v^2 \right]\,,
\end{eqnarray}
where the conformal time $\tau$ is related to the physical time $t$ via $ad\tau = c_Sdt$ and each perturbation mode crosses the sound horizon when $k^2=z_S''/z_S\sim1/\tau^2$. The effective sounds speed $c_S^2$ is expressed as 
\begin{eqnarray}
c_S^2 \equiv \frac{F_S}{G_S} = \frac{G_T^2\Theta H+2G_T\dot{G_T}\Theta-G_T^2\dot{\Theta}-F_T \Theta^2}{G_T\left(G_T \Sigma+3\Theta^2 \right)}\,.
\end{eqnarray}
In terms of the slow-roll parameters, we obtain
\begin{eqnarray}
c_S^2&\simeq& 1+\frac{3-\gamma\left(15+14\gamma \right)}{9\left(1+\gamma\right)^2}\epsilon_1 
\,,
\end{eqnarray}
where only leading order contribution is collected.
By employing the canonical quantization 
\begin{eqnarray}
\hat{v}(\tau,\vec{x}) = \int \frac{d^3k}{(2\pi)^3}\left[v_{k}(\tau)\hat{a}_k e^{i\vec{k}\vec{x}}+v^\ast_{k}(\tau)\hat{a}^\dagger e^{-i\vec{k}\vec{x}} \right]\,,
\end{eqnarray} 
we arrive at a equation for the scalar perturbation modes
\begin{eqnarray}\label{eq:SmodeEq}
v_{k}''+\left(c_S^2 k^2-\frac{\mu_S^2-1/4}{\tau^2}\right)v_{k}=0\,,
\end{eqnarray}
where
\begin{eqnarray}
\frac{z''_{S}}{z_S}&\simeq& \frac{\mu_S^2-1/4}{\tau^2}\,,\qquad \mu_S 
\simeq\frac{3}{2}+\frac{4\epsilon_V-\eta_V}{1+\mathcal{A}}-\frac{\epsilon_V}{(1+\mathcal{A})^2}\,.
\end{eqnarray}
After adopting the Bunch-Davies vacuum for the initial fluctuation modes, the solution to Eq.~(\ref{eq:SmodeEq}) is given by
\begin{eqnarray}\label{eq:Scalsol}
v_k = 2^{\mu_S-\frac{3}{2}}\frac{\Gamma(\mu_S)}{\Gamma(3/2)}\frac{e^{i\left(\mu_S-\frac{1}{2}\right)\frac{\pi}{2}}}{\sqrt{2c_S k}}\left(-c_S k \tau \right)^{\frac{1}{2}-\mu_S}\,.
\end{eqnarray}
The power spectra of the scalar modes can be computed with Eq.~(\ref{eq:Scalsol}) on the large scale $c_S k |\tau|\ll 1$ as
\begin{eqnarray}\label{eq:Ps}
\mathcal{P}_S=\frac{k^3}{2\pi^2}\left|\frac{v_k}{z_S}\right|^2 \simeq \frac{\kappa^2 H^2}{8\pi^2 c_S^3 \epsilon_V}\left(1+\mathcal{A}\right)\,.
\end{eqnarray}
The spectral index of the scalar perturbation modes at the time of horizon crossing is obtained as
\begin{eqnarray}\label{eq:nsindex}
n_S-1&=&\left.\frac{\ln \mathcal{P}_S}{\ln k}\right|_{c_T k = a H}=3-2\mu_S
\simeq \frac{1}{1+\mathcal{A}}\left[2\eta_V-2\epsilon_V\left(4-\frac{1}{1+\mathcal{A}}\right)\right]\,.
\end{eqnarray}
Consequently, the tensor-to-scalar ratio becomes
\begin{eqnarray}\label{eq:rratio}
r=\left.\frac{\mathcal{P}_T}{\mathcal{P}_S}\right.\simeq\frac{16}{1+\mathcal{A}}\epsilon_V\,.
\end{eqnarray}
One can notice here that quantities $n_S$, $n_T$, and $r$ are suppressed by a factor of $(1+\mathcal{A})$, which is not surprising because such suppression was previously discussed in Refs.~\cite{Germani:2010gm, Tsujikawa:2012mk, Yang:2015zgh, Sato:2017qau}. However, what we found as a result of our computation is the additional factor $(1+1/\gamma)$ of suppression in the $\mathcal{A}\gg1$ limit, where $\mathcal{A}\equiv-3c_1(1+1/\gamma)H^2$. Therefore, the presence of the derivative self-interaction and the non-minimal derivative coupling to gravity terms is responsible for this additional suppression factor. The results of standard slow-roll inflation with Einstein gravity $n_S-1=2\eta_V-6\epsilon_V$, $n_T=-2\epsilon_V$, and $r=16\epsilon_V$ is recovered in the $|\mathcal{A}|\ll 1$ limit.  

To be consistent with other related works~\cite{Germani:2010gm, Tsujikawa:2012mk, Yang:2015zgh, Sato:2017qau, Amendola:1993uh, Kamada:2010qe, Kobayashi:2011pc}, we set $c_1=-1/M^2$ where $M$ is a mass scale (see Ref.~\cite{Amendola:1993uh} for the original work) satisfying the quantum gravity constraint $H^2\ll M_p^2$ and $M\kappa\ll 1$~\cite{Germani:2011ua}  and $c_3=-1$ in the following section. Consequently, the high-friction limit can also be rewritten as $H^2\gg M^2$. 

\section{Observational constraints on explicit models}\label{sec:ObsCons}
In the presence of both the non-minimal derivative coupling to gravity and the inflaton derivative self-interaction, we put observational constraints on (A) chaotic inflation~\cite{Linde:1981mu, Kawasaki:2000yn} and (B) natural inflation~\cite{Freese:1990rb,  Germani:2011ua} in this section. In the framework of standard single-field inflationary models with Einstein gravity, these models are disfavored by the current \emph{Planck 2018} plus \emph{BK14} data~\cite{Akrami:2018odb} due to their predictions of large tensor-to-scalar ratio. However, we showed in the previous section that the tensor-to-scalar ratio is significantly suppressed for our model. Thus, based on the information of the $n_S$ and $r$ given in Eqs.~(\ref{eq:nsindex}) and~(\ref{eq:rratio}), we examine the observational bounds of each model in the following subsections. 

\subsection{Chaotic inflation}\label{subsec:chaotInf}
The scalar-field potential for the chaotic inflation model~\cite{Linde:1981mu, Kawasaki:2000yn} is given by
\begin{equation}\label{eq:ChaoticP}
V(\phi)=\frac{\lambda}{n \kappa^4} \left( \kappa \phi\right)^n\,,
\end{equation}
where $\lambda=m^2 \kappa^2$ for a quadratic $n=2$ potential. A shape of the coupling function $\xi(\phi)$ during inflation can be determined from Eqs.~(\ref{eq:Addcond}) with a use of above potential. 
From Eq.~(\ref{eq:Nofefold}), the number of $e$-folds becomes 
\begin{equation}\label{eq:Nefolding}
N=\frac{(\kappa \phi)^2}{2n^2}\left[n+\frac{2\delta}{n+2}(\kappa\phi)^n \right] - \frac{(\kappa \phi_e)^2}{2n^2}\left[n+\frac{2\delta}{n+2}(\kappa\phi_e)^n \right]\,,
\end{equation}
where 
\begin{equation}\label{eq:defdelta}
\delta\equiv\frac{\lambda}{\kappa^2 M^2}\left(1+\frac{1}{\gamma}\right)\,.
\end{equation} 
The scalar field value at the end of inflation $\phi_e$ can be estimated by solving $\epsilon(\phi_e)=1$. In our case, we get
\begin{eqnarray}\label{eq:phiend}
2(\kappa \phi_e)^2\left[n+\delta(\kappa\phi_e)^n \right]=n^3\,.
\end{eqnarray}
From Eqs.~(\ref{eq:nsindex})--(\ref{eq:rratio}), we obtain the spectral index and the tensor-to-scalar ratio as
\begin{eqnarray}\label{eq:nsrphi}
n_S = 1- \frac{n^2\left[n(n+2)+2(n+1){\delta}\left(\kappa \phi\right)^n\right]}{(\kappa \phi)^2\left[n+{\delta}\left(\kappa \phi\right)^n\right]^2}\,,\qquad
r = \frac{8n^3}{\left(\kappa \phi\right)^2\left[n+{\delta}(\kappa \phi)^n\right]}\,.
\end{eqnarray}

In order to examine the observational consistency of the model, we often express $n_S$ and $r$ as fuctions of $N$. For that purpose, we solve Eqs.~(\ref{eq:Nefolding}) and (\ref{eq:phiend}) for $\phi$ and $\phi_e$. Let us assume for computational simplicity that the ${\delta}\ll 1$ is a small parameter during inflation and then expand $\phi$ to the leading order in $\delta$ as follows
\begin{equation}\label{eq:phiexpnd}
\phi = \phi^{(0)}+{\delta}\phi^{(1)}+\mathcal{O}({\delta}^2)\,.
\end{equation} 
Substituting Eq.~(\ref{eq:phiexpnd}) into Eqs.~(\ref{eq:Nefolding}) and (\ref{eq:phiend}), we obtain
\begin{eqnarray}\label{eq:phieandphi}
\kappa\phi_e&=&\left(\frac{n^2}{2}\right)^{\frac{1}{2}}\left[1-\frac{\delta}{2n}\left(\frac{n^2}{2}\right)^\frac{n}{2}\right]\,,\qquad 
\kappa\phi = \left(2 n \tilde{N}\right)^{\frac{1}{2}}\left[ 1-\frac{{\delta}}{n(n+2)}\left(2 n \tilde{N}\right)^{\frac{n}{2}}\right]\,,
\end{eqnarray}
where 
\begin{equation}
\tilde{N}=N+\frac{n}{4}\left[1+\frac{{\delta}}{n^2}\left(\frac{n^2}{2}\right)^{\frac{n}{2}}\right]\,.
\end{equation}
After plugging Eq.~(\ref{eq:phieandphi}) into Eq.~(\ref{eq:nsrphi}), we finally express the observable quantities in terms of $N$ as follows
\begin{eqnarray}
n_S=1-\frac{n+2}{2\tilde{N}}\,,\qquad r = \frac{4n}{\tilde{N}}\left[1+\frac{{\delta}}{n+2}\left(2n\tilde{N}\right)^{\frac{n}{2}} \right]^{-1}\,.
\end{eqnarray}

\begin{figure}[h!]
\includegraphics[width=0.7\textwidth]{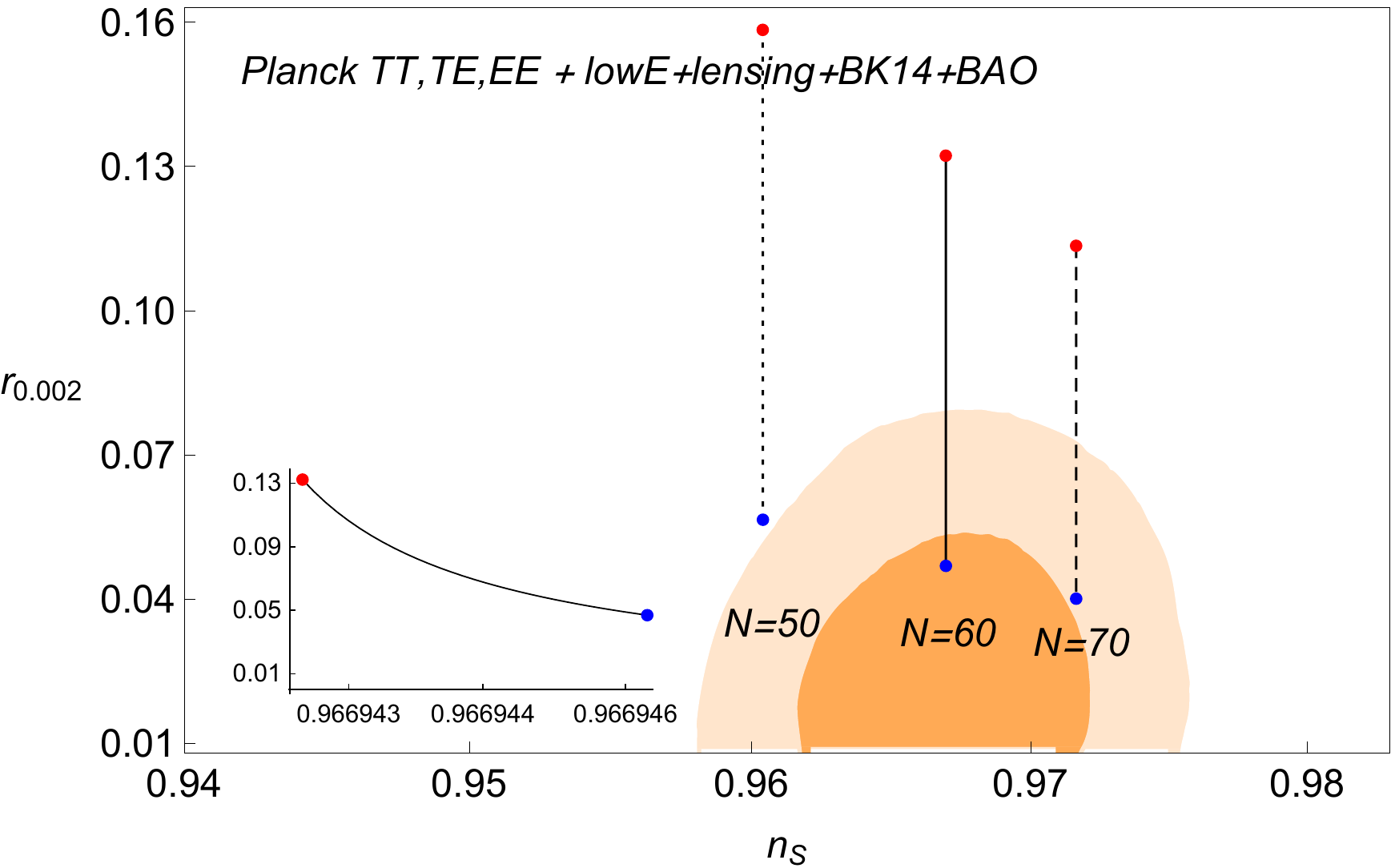}
\caption{{\small Observational constraints on chaotic inflation with $n=2$ in Eq.~(\ref{eq:ChaoticP}). The parameter increases from the red to blue points and ranges between $0\leq \delta < \delta_{max}$, where $\delta_{max}\simeq 3.5727\times10^{-2}$, $3.0094\times10^{-2}$, and $2.6008\times10^{-2}$ for $N=50$ (dotted), $60$ (solid), and $70$ (dashed) lines, respectively. The background shaded regions indicate the $1\sigma$(darker) and $2\sigma$(lighter) contours of the observational data~\cite{Akrami:2018odb}. The inset plot shows a closer look on the $N=60$ solid line. 
}} \label{fig:Chaoticinf}
\end{figure}

In Fig.~\ref{fig:Chaoticinf}, together with the observational data, we present the theoretical predictions of chaotic inflation with the quadratic potential. The background shaded regions show the $1\sigma$(darker orange) and $2\sigma$(lighter orange) contours of the observational data by \emph{Planck TT, TE, EE $+$ lowE $+$ lensing $+$ BK14 $+$ BAO}~\cite{Akrami:2018odb}. The red points indicate the $\delta\rightarrow 0$ limit or correspond to results of the standard single-field chaotic inflation models with Einstein gravity: for example, we have $(n_S,r)=(0.966942,0.1322)$ at the $N=60$ red point. 
For the large--field scenario of chaotic inflation, the scalar field value in the beginning of inflation is assumed to be larger than its value at the end of inflation \emph{i.e.,} $\phi>\phi_e $. On the basis of this criteria, we find from Eq.~(\ref{eq:phieandphi}) that the $\delta$ cannot be larger than a certain value or there exists a $\delta_{max}$. For $n=2$ in Eq.~(\ref{eq:ChaoticP}), we calculated the numerical values: $\delta_{max}\simeq 3.5727\times10^{-2}$, $3.0094\times10^{-2}$, and $2.6008\times10^{-2}$ for $N=50$, $60$, and $70$, respectively. These maximum values give us the blue points in Fig.~\ref{fig:Chaoticinf}. For example, the $N=60$ blue point at $(n_S, r)\simeq(0.966946, 0.0468)$ corresponds to the $\delta_{max}=3.0094\times10^{-2}$ value. Thus, the $\delta$ varies between zero and $\delta_{max}$. In the figure, the $\delta$ increases from red to blue points and significantly reduces the $r$ value; its impact on $n_S$ is a slight increment, see the plot embedded in Fig.~\ref{fig:Chaoticinf}.

For the $n=4$ potential, we also obtained  $\delta_{max}\simeq 1.24\times 10^{-4}$, $8.7884\times10^{-5}$, and $6.5565\times10^{-5}$ for $N=50$, $60$, and $70$, respectively, and estimated predictions ($n_S, r)\simeq (0.9511, 0.0584)$ if $N=60$ and $(n_S, r) \simeq (0.9629, 0.0869)$ if $N=80$. However, these predictions were not presented in Fig.~\ref{fig:Chaoticinf} as they would reside outside of the 2$\sigma$ contour hence disfavored by the current data~\cite{Akrami:2018odb}. This is why we plot only $n=2$ potential in Fig.~\ref{fig:Chaoticinf}.

As can be seen in Fig.~\ref{fig:Chaoticinf}, chaotic inflation with the quadratic potential is consistent with the current observational data for $N=60$ and $N=70$ $e$-folds, whereas $N=50$ $e$-folds appears to be disfavored because its prediction residing outside of the $2\sigma$ contour. After taking $r_{0.002}<0.065$ into account from Ref.~\cite{Akrami:2018odb} ($95\%$ CL by \emph{Planck TT, TE, EE $+$ lowE $+$ lensing $+$ BK14 $+$ BAO}), a lower limit of the $\delta$ can be tightly constrained. For $N=60$ $e$-folds, for example, the favored $\delta$ range by observation is given as: $1.7618\times10^{-2}\lesssim \delta_{obs} < 3.0094\times10^{-2}$. 

Using the $\delta$ values favored by observation together with the normalized value for the primordial scalar perturbation $\mathcal{P}_S\simeq 2.09\times 10^{-9}$ for $k_\ast=0.05\text{Mpc}^{-1}$~\cite{Akrami:2018odb}, we can find the relation between the $\gamma$ and $M$ parameters from Eq.~(\ref{eq:Ps}). By substituting Eq.~(\ref{eq:ChaoticP}) into Eq.~(\ref{eq:Ps}), we obtain
\begin{eqnarray}\label{eq:Pschaot}
\frac{\delta_{obs}}{12n^4 \pi^2}\left(\frac{\kappa^2 M^2}{1+1/\gamma}\right)(\kappa\phi)_{60}^{n+2}\left[n+\delta_{obs} (\kappa\phi)_{60}^n\right]\simeq 2.09\times10^{-9}\,,
\end{eqnarray}
where $\delta_{obs}$ is the numerical value of $\delta$ in the range favored by the observational data and $(\kappa\phi)_{60}$ is the value of $\kappa\phi$ at $N=60$. After putting Eq.~(\ref{eq:phieandphi}) into Eq.~(\ref{eq:Pschaot}) and setting the $n$ value, we basically have two parameters: $M$ and $\gamma$, to tune the correct amplitude of the primordial scalar perturbation. 

\begin{figure}[h!]
\centering
{\includegraphics[width=0.7\textwidth]{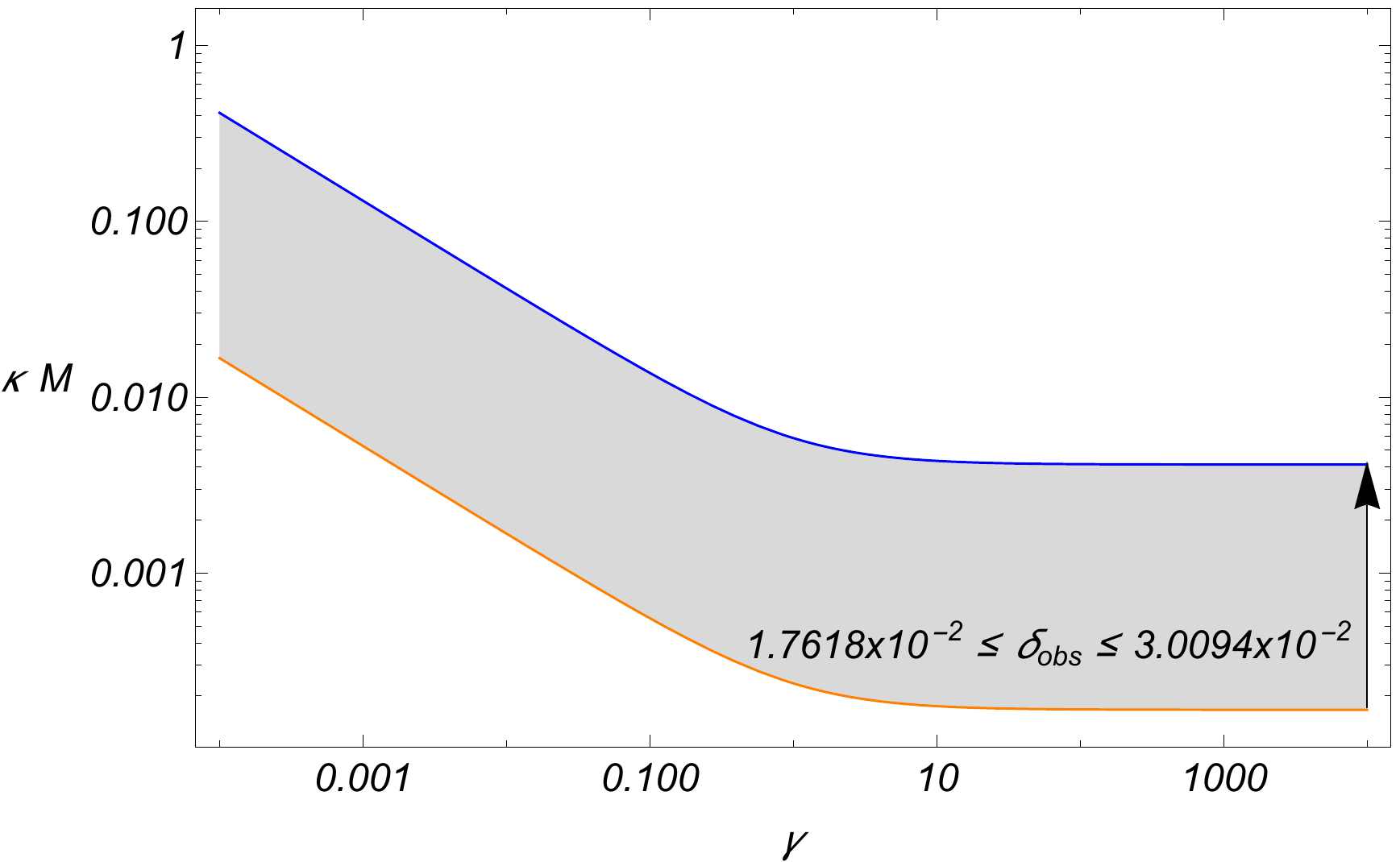}}
\caption{\small The relation between $\gamma$ and $M$, where $n=2$ and $N=60$. The $\delta$ increases along the direction of the arrow.
}\label{fig:kappaMofgamma}
\end{figure}

Fig.~\ref{fig:kappaMofgamma} shows the relation between $\gamma$ and $M$ from Eq.~(\ref{eq:Pschaot}) for the quadratic potential that gives the correct amplitude of the scalar power spectrum $\mathcal{P}_S$. The range of $\delta_{obs}$ value is $1.7618\times10^{-2}\lesssim \delta_{obs} < 3.0094\times10^{-2}$, increasing from the orange to the blue line. The potential parameter $\lambda$ can be estimated for a given set of $\gamma$ and $M$. One can show by using Eqs.~(\ref{eq:defdelta}) and~(\ref{eq:Pschaot}) that $\lambda=const.$ along each of the orange and blue line. Taking the numerical values of $\gamma$ and $M$ by matching on the figure and using Eq.~(\ref{eq:defdelta}), we find for our model that $4.8513\times10^{-10}\lesssim \lambda=m^2 \kappa^2\lesssim 5.1109\times10^{-7}$ is favored by the data~\cite{Akrami:2018odb}.

Furthermore, Fig.~\ref{fig:kappaMofgamma} shows that the scale $M$ approaches to the Planck scale $M_p=\kappa^{-1}$ in the $\gamma\ll1$ regime. This regime corresponds to the case where the effects of the derivative self-interaction $G_3(\phi, X)$ becomes stronger than that of the non-minimal derivative coupling to gravity $G_5(\phi, X)$ during inflation. If we impose $\kappa M\ll 1$ to avoid the quantum gravity, $\gamma\gtrsim 10^{-4}$ seems to be more favored. Thus, too small values ( as small as $\gamma<10^{-4}$) may violate the quantum gravity constraint, see Ref.~\cite{Germani:2011ua} for further details. The scale $M$ is much smaller than the Planck scale in both the $\gamma\sim1$ and $\gamma\gg1$ regimes and evolves differently in each regime. As can be seen in the figure, the evolution is $\kappa M\sim (1+1/\gamma)$ in the $\gamma \sim 1$ regime and nearly a constant for the $\gamma\gg1$ regime, where the $G_5(\phi, X)$ term plays in an important role during inflation. 

Thus, for chaotic inflation with the quadratic potential, we emphasize that the $\gamma\ll1$ limit or the G-inflation scenario, may suffer from the quantum gravity constraints, whereas no such issues are evident in the $\gamma\sim \mathcal{O}(1)$ regime or for inflationary models with the non-minimal derivative coupling to gravity.
 
\subsection{Natural inflation}
The scalar-field potential for natural inflation is given by~\cite{Freese:1990rb, Germani:2011ua}
\begin{eqnarray}\label{eq:NatPot}
V(\phi) = \Lambda^4\left[ 1+\cos\left(\frac{\phi}{f}\right)\right]\,,
\end{eqnarray}
where $\Lambda$ and $f$ are model parameters having dimension of mass.~\footnote{According to Ref.~\cite{Germani:2011ua}, $f\kappa\ll1$ and $\Lambda \kappa\ll1$ are assumed in order to avoid trans-Planckian masses and satisfy the quantum gravity constraint such that the curvature should be smaller than the Planck scale, respectively.} For the given potential, the coupling function $\xi(\phi)$ during inflation can be determined from Eq.~(\ref{eq:Addcond}). 
The number of $e$-folds is obtained as
\begin{eqnarray}\label{eq:Nofefoldphi}
N 
&=&-\left.\alpha \left[\cos\left(\frac{\phi}{f}\right)+4\ln\left(\sin\left(\frac{\phi}{2 f}\right)\right)\right]\right|_{\phi_e}^\phi
\,,
\end{eqnarray}
where 
\begin{equation}
\alpha=\left(1+\frac{1}{\gamma}\right)\frac{f^2 \kappa^4\Lambda^4}{M^2}\,.
\end{equation}
By solving $\epsilon_1(\phi_e)=1$, we find the field value at the end of inflation 
\begin{eqnarray}\label{eq:cosphie}
\cos\left(\frac{\phi_e}{f}\right)=\frac{\sqrt{1+16\alpha}-1}{4\alpha}-1\,.
\end{eqnarray}

\begin{figure}[h!]
\centering
\includegraphics[width=0.7\textwidth]{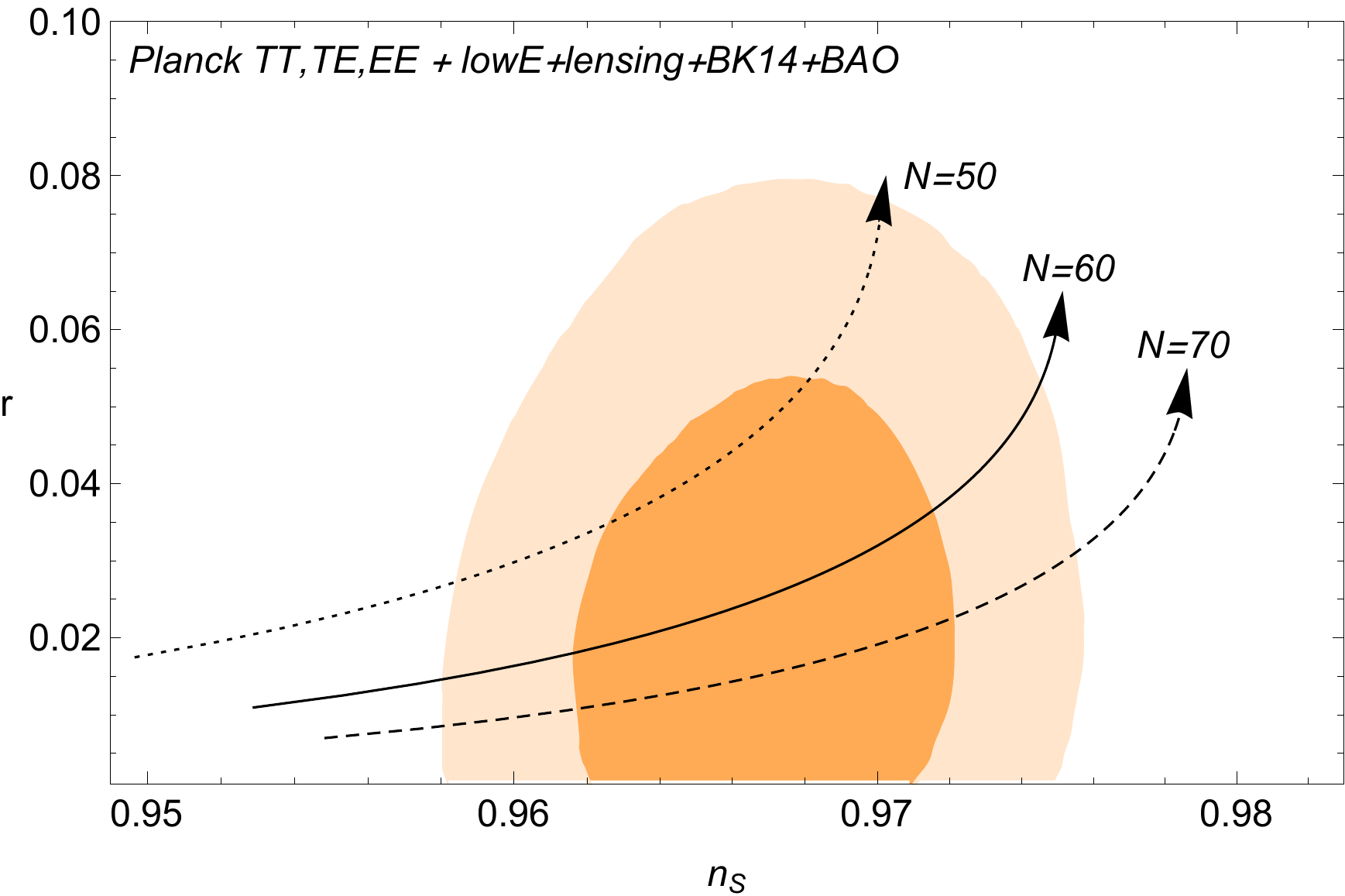} 
\caption{\small Observational constraints on natural inflation. The parameter ranges between $12\leq \alpha \leq 10^3$ for each dotted ($N=50$), solid ($N=60$), and dashed ($N=70$) line. The arrows indicate the increasing direction of $\alpha$.}\label{fig:constraint}
\end{figure}

\begin{figure}[h!]
\centering
\includegraphics[width=0.7\textwidth]{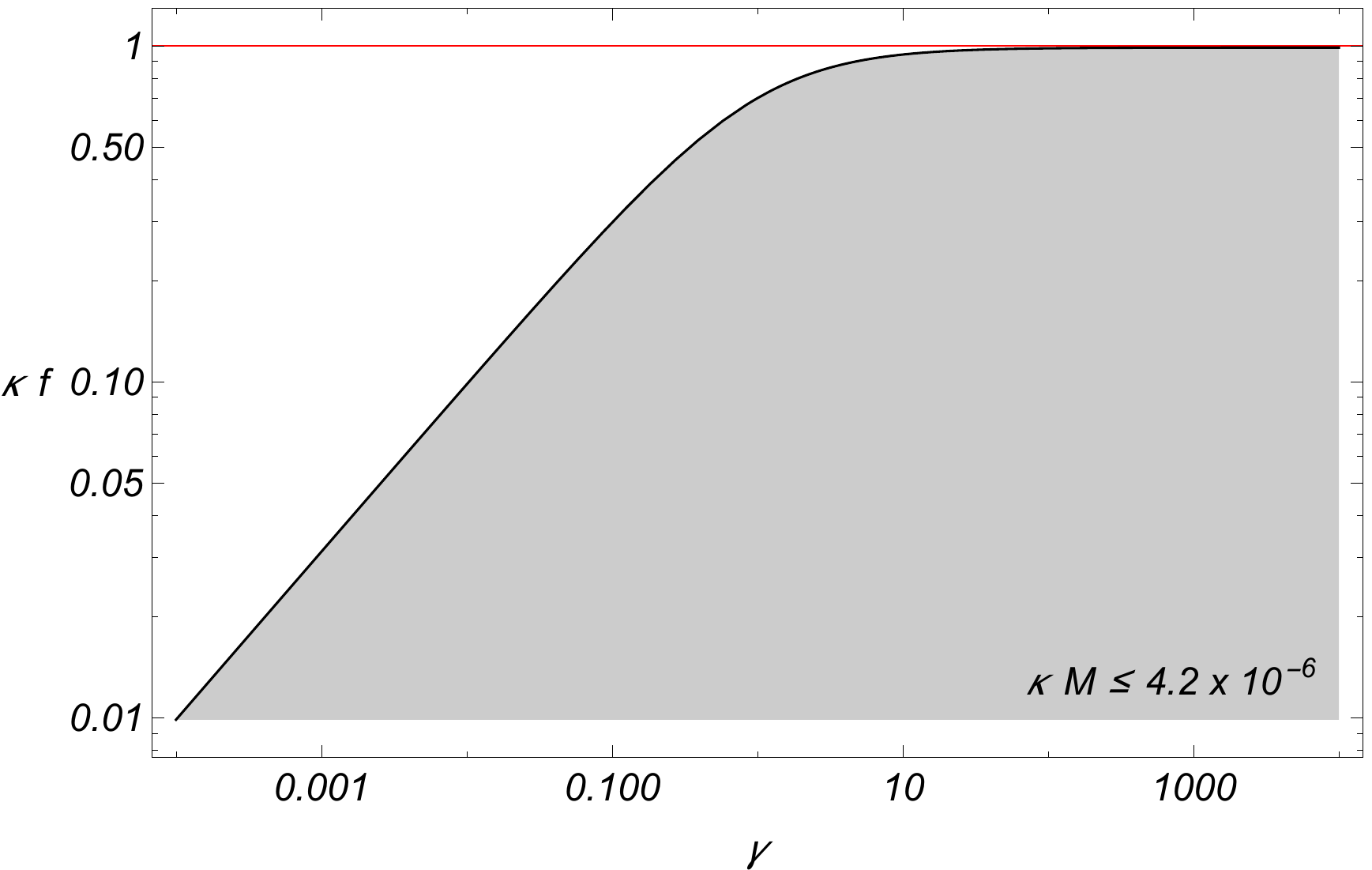}
\caption{\small The parameter space from Eq.~(\ref{eq:Psnat}) where $\kappa \Lambda\simeq4.2418\times10^{-3}$. For the shaded region $\kappa M\leq4.2\times10^{-6}$. }\label{fig:kappaFofgamma}
\end{figure}

From Eqs.~(\ref{eq:nsindex})--(\ref{eq:rratio}), for the potential in Eq.~(\ref{eq:NatPot}), the observable quantities are calculated 
\begin{eqnarray}\label{eq:nsandr}
n_S 
    = 1- \frac{2\left[2-\cos\left(\phi/f\right)\right]}{\alpha \left[1+\cos(\phi/f)\right]^2} \,, \qquad
 r  = 
 \frac{8\left[1-\cos(\phi/f)\right]}{\alpha\left[1+\cos(\phi/f)\right]^2}\,.
\end{eqnarray}
Here, we can express $\cos\left(\phi/f\right)$ in terms of $N$ using Eq.~(\ref{eq:Nofefoldphi}). In the $\gamma\gg1$ limit, Eq.~(\ref{eq:nsandr}) reproduces the results of Ref.~\cite{Tsujikawa:2012mk}. In Fig.~\ref{fig:constraint}, we plot predictions of natural inflation and observational constraints in the $n_S$ -- $r$ plane, where the range of $\alpha$ increases from the lower end to the upper end within the range $12\leq \alpha \leq 10^3$. The dotted, solid, and dashed lines correspond to $N=50$, $60$, and $70$, respectively. The background shaded regions are the same as Fig.~\ref{fig:Chaoticinf}. The direction of arrow indicates that both the $n_S$ and $r$ increases as the $\alpha$ increases. 
According to the figure, the theoretical prediction of natural inflation is found to be consistent with the current observation for the certain range of $\alpha$. The range $\alpha$ is constrained to be $26.004\lesssim\alpha_{obs}\lesssim 193.92$ for $N=50$, $18.57\lesssim\alpha_{obs}\lesssim 37.79$ for $N=60$, and $16.23\lesssim\alpha_{obs}\lesssim 25.66$ for $N=70$, respectively, after taking $n_S=0.9670\pm0.0037$ and $r_{0.002}\leq0.065$ from \emph{Planck TT, TE, EE $+$ lowE $+$ lensing $+$ BK14 $+$ BAO}~\cite{Akrami:2018odb} into account.
Using Eqs.~(\ref{eq:NatPot}) and~(\ref{eq:Ps}), the CMB normalization by~\cite{Akrami:2018odb} at $N=60$ corresponds to
\begin{eqnarray}\label{eq:Psnat}
\frac{\alpha_{obs}}{12\pi^2}\frac{\,\,\left(1+x_{60}\right)^3}{\left(1-x_{60}\right)}  \kappa^4\Lambda^4 \simeq 2.1\times 10^{-9}\,,
\end{eqnarray}
where $x_{60}=\cos(\phi_{60}/f)$ and $\alpha_{obs}$ is chosen to take values between $18.57\lesssim \alpha_{obs}\lesssim37.79$ at 68\% CL. From Eq.~(\ref{eq:Psnat}), the values of $\Lambda$ corresponding to the CMB normalization are given in the following range 
\begin{eqnarray}
4.2822\times 10^{-3} \lesssim \kappa \Lambda \lesssim 5.0631\times 10^{-3}\,.
\end{eqnarray}
Fig.~\ref{fig:kappaFofgamma} shows the parameter space that gives the correct amplitude of the scalar power spectrum $\mathcal{P}_S$. If we impose $\kappa f\ll1$ to avoid the trans-Planckian masses, the \emph{Planck 2018} result at 68\% CL~\cite{Akrami:2018odb} leads to the somewhat tighter constraint $\kappa M \leq 4.2\times10^{-6}$ on the mass scale. 
As the $M$ increases toward the black curve where $\kappa M = 4.2 \times 10^{-6}$, the figure also shows that the $f$ approaches to the Planck scale $\kappa f \simeq 1$ (the horizontal red line) in the $\gamma\gg1$ regime. Thus, for the too large values of $\gamma$ as large as $\gamma\gtrsim10^4$, inflationary models with non-minimal derivative coupling to gravity may not be able to avoid from the trans-Planckian masses by having $f\gtrsim M_{pl}$.  In the $\gamma\lesssim1$ regime, on the other hand, the model is not only consistent with the observational data but also respects the $\kappa f\ll1$ and $\kappa M\ll 1$ constraints. 

{\section{Conclusion}\label{sec:conc}}
We have studied inflationary models with the non-minimal derivative coupling to gravity and the derivative self-interaction of the scalar field. After deriving the background equations of motion, we introduced the special relation Eq.~(\ref{eq:Addcond}) that holds during inflation. If we employ the relation, our model approaches to the single term limit with respect to the model parameters. Thus, we did not need to specify the coupling function $\xi(\phi)$ in the present study. In addition, the contributions of each term can be conveniently controlled by the $\gamma$ parameter in Eq.~(\ref{eq:Addcond}), which reflects the strength of the inflaton derivative self-interactions with respect to the non-minimal derivative coupling to gravity.  

The observable quantities including the power spectra for scalar and tensor perturbation modes, the spectral indices, and the tensor-to-scalar ratio are obtained in Eqs.~(\ref{eq:Pt}), (\ref{eq:ntindex}), (\ref{eq:Ps}), (\ref{eq:nsindex}), and (\ref{eq:rratio}). We found as a result of our analytic computation that the observable quantities are suppressed by a factor of ($1+\mathcal{A}$), or $(1+1/\gamma)$ in the $\mathcal{A}\gg 1$ limit. The suppression is mainly due to the presence of both the non-minimal derivative coupling to gravity and the inflaton derivative self-interaction terms. 

We then placed observational constraints on the chaotic and natural inflation models using their theoretical predictions for the $n_S$ and $r$. Figs.~\ref{fig:Chaoticinf} and~\ref{fig:constraint} show that, for certain ranges of the model parameters, the both models are consistent with the current observational data~\cite{Akrami:2018odb} mainly due to the suppressed tensor-to-scalar ratio. In Figs.~\ref{fig:kappaMofgamma} and~\ref{fig:kappaFofgamma}, the shaded regions illustrate the parameter spaces that give the correct amplitude of the scalar power spectrum $\mathcal{P}_S$ for each inflation model and satisfy the associated $f\ll M_{pl}$ and $M\ll M_{pl}$ constraints. Although a broad range of the $\gamma$ parameter is supported by the observational data~\cite{Akrami:2018odb}, the values that are as small as $\gamma\lesssim10^{-4}$ for chaotic inflation and as large as $\gamma\gtrsim10^{4}$ for natural inflation may suffer from avoiding the trans-Planckian mass and the quantum gravity constraints. There is no such issues apparent in the $\gamma\sim\mathcal{O}(1)$ regime, where both the derivative coupling and self-interaction terms play equally important role during inflation, and both inflationary models fit well to the observational data~\cite{Akrami:2018odb}.

The \emph{Planck 2018} result leads to the somewhat tighter constraint on the potential parameters: $4.8513\times10^{-10}\lesssim \lambda=m^2 \kappa^2\lesssim 5.1109\times10^{-7}$ for chaotic inflation with the quadratic potential and $4.2822\times 10^{-3} \lesssim \kappa \Lambda \lesssim 5.0631\times 10^{-3}$ for natural inflation with the cosine potential. In addition, the observational bounds on the mass scale for natural inflation is constrained to be $\kappa M \lesssim 4.2\times10^{-6}$ at 68\% CL; therefore, the scale $f$ is smaller than the Planck scale $f \lesssim M_{pl}$. 

Based on our finding, we consider the background dynamics of the system including the post-inflationary evolution need to be further analyzed for the better understanding of the system Eq.~(\ref{eq:NMDCandGinf}). In addition, one would expect the introduction of new terms and a new mass scale $M$ may produce non-Gaussian fluctuations larger than those in GR even for the same $V(\phi)$. We leave these as our future extensions to our present study.\\

\begin{acknowledgments}
We thank Ryusuke Jinno, Masahide Yamaguchi for their constructive comments in the present analysis as well as on the manuscript. We also thank Kohei Kamada for reading and providing valuable comments on the manuscript. This work was supported by IBS under the project code IBS-R018-D1. 
\end{acknowledgments}


\end{document}